\documentclass[10pt]{article}

\usepackage{graphicx}
\usepackage{graphics}
\usepackage{amscd}
\usepackage{amsmath}
\usepackage{amsfonts}
\usepackage{amssymb}
\usepackage{cite}
\usepackage{color}

\DeclareMathOperator{\wt}{wt_H}

\newcommand{\bF}{ {\mathbb F}}

\newtheorem{theorem}{Theorem}

\newtheorem{corollary}[theorem]{Corollary}

\newtheorem{example}[theorem]{Example}

\newtheorem{lemma}[theorem]{Lemma}

\newtheorem{remark}[theorem]{Remark}

\begin{document}
\title{ Subfield codes of linear codes from perfect nonlinear functions\\ and their duals\footnote{
 *Corresponding author.\,\, E-Mail addresses:
 dzheng@hubu.edu.cn(D.\ Zheng), waxiqq@163.com (X.\ Wang)
 liyayao2020@163.com(Y.\ Li), yuanmu847566@outlook.com (M.\ Yuan)}
 }

\author{  Dabin Zheng*, Xiaoqiang Wang, Yayao Li, Mu Yuan}

\date{ \small Hubei Key Laboratory of Applied Mathematics, \\
Faculty of Mathematics and Statistics, Hubei University, Wuhan 430062, China
}

\maketitle

\leftskip 1.0in
\rightskip 1.0in
\noindent {\bf Abstract.} Let $\mathbb{F}_{p^m}$ be a finite field with $p^m$ elements, where $p$ is an odd prime and $m$ is a positive integer. Recently, \cite{Hengar} and \cite{Wang2020} determined the weight distributions of subfield codes with the form
 $$\mathcal{C}_f=\left\{\left(\left( {\rm Tr}_1^m(a f(x)+bx)+c\right)_{x \in \mathbb{F}_{p^m}}, {\rm Tr}_1^m(a)\right)\, : \, a,b \in \mathbb{F}_{p^m}, c \in \mathbb{F}_p\right\}$$
for $f(x)=x^2$ and $f(x)=x^{p^k+1}$, respectively, where $k$ is a nonnegative integer. In this paper, we further investigate the subfield code $\mathcal{C}_f$ for $f(x)$ being a known perfect nonlinear function over $\mathbb{F}_{p^m}$ and generalize some results in \cite{Hengar,Wang2020}.
 The weight distributions of the constructed codes are determined by applying the theory of quadratic forms and the properties of perfect nonlinear functions over finite fields. In addition, the parameters of the duals of these
codes are also determined. Several examples show that some of our codes and their duals have the best known parameters with respect to the code tables  in \cite{MGrassl}. The duals of some proposed codes are optimal with respect to the Sphere Packing bound if $p\geq 5$.

\vskip 6pt

\vskip 35pt

\leftskip 0.0in
\rightskip 0.0in
\section{Introduction}
Let $p$ be an odd prime and $ \mathbb{F}_{p^m}$ be a finite field of size $p^m$. An $[n,k,d]$ code $\mathcal{C}$ over the finite field $\mathbb{F}_{p^m}$ is a $k$-dimensional linear subspace of $\mathbb{F}_{p^m}^{n}$ with  minimum Hamming distance $d$. An $[n,k,d]$ code is called {\it distance-optimal} if there dose not exist $[n,k,d+1]$ code \cite{Dingar}. The Hamming weight of a codeword $\boldsymbol{c}=(c_{0},c_{1},\cdots,c_{n-1})\in\mathcal{C}$ is the number of nonzero $c_{i}$ for $0\le i\le n-1$. Let $A_{i}$ denote the number of nonzero codewords with Hamming weight $i$ in $\mathcal{C}$. The {\it weight enumerator} of $\mathcal{C}$ is defined by $1+A_{1}x+A_{2}x^2+\cdots+A_{n}x^n$ and the sequence $(1,A_{1},\cdots,A_{n})$ is called the {\it weight distribution} of $\mathcal{C}$. The weight distribution of a code is used to estimate the error correcting capability and compute the error probability of error detection and correction of the code \cite{Klove2007}. The weight distributions of linear codes have also application in cryptography and combinatorics. Hence, the research of the weight distribution of a linear code is a hot topic in coding theory. The recent progress on weight distributions of linear codes can be seen in~\cite{Ding2015, Ding2016, HengYue20162, HengYue2017,  Lietal2016, LuoCaoetal2018, Tan2018, Tang2018, TangLietal2016, Wangetal2016, Xiaetal2017, ZhouDing2013, ZhouDing2014, ZhouLietal2015} and the references therein.

Let $f(x)$ be a function from $\bF_{p^m}$ to itself, then $f(x)$ is called a perfect nonlinear (PN) function or planar function if
$$\max_{a\in \mathbb{F}_{p^m}^*}\max_{b\in \mathbb{F}_{p^m}}\vert \{ x \in \mathbb{F}_{p^m}: f(x+a)-f(x)= b \} \vert= 1.$$
PN functions were first introduced to construct finite projective planes by Dembowski and Ostrom \cite{Dembowski1968} in 1968.
Then looking for new non-equivalent PN functions aroused a lot of interest for many researchers in cryptography since these functions are optimally resistant to linear and differential cryptanalysis when used in DES-like cryptosystems.
Up to now, all known PN functions from $\bF_{p^m}$ to $\bF_{p^m}$ with explicit expressions are equivalent to one of the following
polynomials~\cite{Bierbrauer2010,Dembowski1968,Ding2006,Zha2009,Coulter1997,Budaghyan2008}:
\begin{description}
\item{$\bullet$} $f_{1}(x)=x^{p^{k}+1}$, where $k \ge 0$ and $ 2\nmid \frac{m}{{\rm gcd}(m,k)}$  (Dembowski and Ostrom \cite{Dembowski1968}).
\item{$\bullet$} $f_{2}(x)=x^{\frac{p^{k}+1}{2}}$, where $p=3$, $ 2\nmid k$, and ${\rm gcd}(m,k)=1$ (Coulter and Matthews \cite{Coulter1997}).
\item{$\bullet$} $f_{3}(x)=x^{10}-\beta x^{6}-\beta ^{2}x^{2}$, where $p=3$, $ 2\nmid m$ and $\beta \in \mathbb{F}_{p^m}^{*}$ (Ding and Yuan \cite{Ding2006}).
\item{$\bullet$} $f_{4}(x)=\beta x^{p^k+1}-\beta^{p^s}x^{p^{ls}+p^{-ls+k}}$, where $\beta$ is a primitive element in $\mathbb{F}_{p^m}$, $m=3s$, ${\rm gcd}(3,s)=1$, $2\nmid \frac{s}{{\rm gcd}(s,k)}$, $k\equiv\pm s \pmod{3}$, $l=1$ if  $s-k=0 \pmod{3}$ and $l=-1$ if  $s+k=0 \pmod{3}$ (Zha, Kyureghyan and Wang \cite{Zha2009}).
\item{$\bullet$} $f_{5}(x)=(\beta x)^{p^k+1}-((\beta x)^{p^k+1})^{p^s}+\sum_{i=0}^{s-1}c_{i}x^{p^i(p^s+1)}$, where $m=2s$, both $s$ and $k$ are positive integers,  ${\rm gcd}(s+k,2s)={\rm gcd}(s+k,s)$, ${\rm gcd}(p^k+1,p^s+1)\ne{\rm gcd}(p^k+1,\frac{p^s+1}{2}),$
    $\beta \in \mathbb{F}_{p^m}^{*}$ and $\sum_{i=0}^{s-1}c_{i}x^{p^i}$ is a permutation polynomial of $\mathbb{F}_{p^m}$ (Budaghyan and Helleseth \cite{Budaghyan2008}).
\item{$\bullet$} $f_{6}(x)=\beta x^{p^s+1}+zx^{p^k+p^t}+z^{p^s}x^{p^{k+s}+p^{s+t}}+\sum_{i=0}^{s-1}w_{i}x^{p^i(p^s+1)}$, where $m=2s$, both $s$ and $k$ are positive integers, $w_{i}  \in \mathbb{F}_{p^s}$, $ 2\nmid \frac{m}{{\rm gcd}(m,t-k)}$, $\beta\in \mathbb{F}_{p^m}\setminus \mathbb{F}_{p^s}$ and $z=\alpha^{r}$ for $\alpha$ being a primitive element in $\mathbb{F}_{p^m}$ and ${\rm gcd}(p^{k-t}+1,p^s+1)\nmid r$ (Budaghyan and Helleseth \cite{Budaghyan2008}).
\item{$\bullet$} $f_{7}(x)=x^{p^t+1}-\beta x^{p^{2s}+p^{s+t}}$, where $m=3s$, $s^\prime=\frac{s}{{\rm gcd}(s,t)}$ is odd, $t^\prime=\frac{t}{{\rm gcd}(s,t)}$, ${\rm ord}(\beta)=p^{2s}+p^s+1$, $t^\prime+s^\prime\equiv0\pmod{3}$ or $p^s\equiv p^t\pmod{3}$ (Bierbrauer \cite{Bierbrauer2010}).
\item{$\bullet$} $f_{8}(x)=x^{p^t+1}-\beta x^{p^{3s}+p^{t+s}}$, where $m=4s$, $2\nmid \frac{2s}{\gcd(2s,t)}$, $p^s\equiv p^t\equiv 1\pmod{4}$, ${\rm ord}(\beta)=p^{3s}+p^{2s}+p^s+1$ (Bierbrauer \cite{Bierbrauer2010}).
\end{description}
Except for the Coulter-Matthews function $f_2(x)$, the other known PN functions have the algebraic degree~$2$. Quadratic homogeneous polynomials are
called DO polynomials. PN functions with the algebraic degree~2 are called PN-DO functions. Then all known PN functions are divided into PN-DO functions and Coulter-Matthews functions.

Subfield codes were first considered in \cite{Canteaut2000} and \cite{Carlet1998} without using the name ``subfield codes". The definition of subfield codes was first given by \cite[p.5117]{Cannon2013} and a Magma function for subfield codes is actual operated. Recently, Ding and Heng in \cite{Dingar} proved some basic results about subfield codes of linear codes and gave their trace representation.
Assume that $\mathbb{F}_{p^m}=\{x_1,x_2,\cdots,x_{p^m}\}$ and $f(x)$ is a polynomial over $\mathbb{F}_{p^m}$. Let $\mathcal{C}$ be a $[p^m+1,3]$ code with the generator matrix
\begin{equation}\label{eq:matrixA'}
\begin{split}
 G = \left(
\begin{array}{cccccc}
    f(x_1) & f(x_2) & \cdots & f(x_{p^m}) & 1  \\
     x_1 & x_2 & \cdots & x_{p^m} & 0  \\
     1 & 1 & \cdots & 1 & 0
\end{array}
\right).
\end{split}
\end{equation}
Ding and Heng in~\cite{Dingar} showed that the subfield code of $\mathcal{C}$ can be represented by the following trace form:
\begin{equation}\label{code0}
\begin{split}
\mathcal{C}_f=\left\{\left(\left( {\rm Tr}_1^m\left(af(x)+bx\right)+c\right)_{x \in \mathbb{F}_{p^m}}, {\rm Tr}_1^m(a)\right) : \, a,b \in \mathbb{F}_{p^m}, c \in \mathbb{F}_p\right\}.
\end{split}
\end{equation}
For a general polynomial $f(x)$, it is very hard to determine the Hamming weights and their corresponding frequencies of codewords in $\mathcal{C}_f$ for $(a,b,c)$ running through
$(\mathbb{F}_{p^m}, \mathbb{F}_{p^m}, \mathbb{F}_p)$. But they obtained some distance-optimal subfield codes with respect to the Sphere Packing bound from some well-known codes \cite{Hengar,Hengar1,Hengar2, Hengar3} (for example, ovoid codes, hyperoval codes, conic codes, arc codes and MDS codes). In particular, Heng
and Ding in~\cite{Hengar} studied the subfield code $\mathcal{C}_f$ with the form (\ref{code0}) for $f(x)=x^2$. Later, Wang et al. \cite{Wang2020} extended their work and studied the subfield code $\mathcal{C}_f$ for $f(x)=x^{p^k+1}$, where $k$ is a nonnegative integer. For the other results about the subfield codes of linear codes, the readers can refer to \cite{Wang2019,Can2020}.

Along the line of the work in \cite{Hengar,Wang2020}, in this paper, we further study the weight distribution of $\mathcal{C}_f$ and the parameters of its dual for $f(x)$ being a known PN function. Firstly, by applying the relation between the type and rank of quadratic forms, we determine the weight distribution of $\mathcal{C}_f$ and the parameters of its dual for $f(x)$ being a known PN-DO function. Secondly, by solving some special equations over $\bF_{3^m}$ and the Pless power-moment identities,  we determine the weight distribution of $\mathcal{C}_f$ and the parameters of its dual for $f(x)$ being a Coulter-Matthews function. Several examples show that some of our codes and their duals have the best known parameters with respect to the code tables  in \cite{MGrassl}. The duals of some proposed codes are optimal with respect to the Sphere Packing bound if $p\geq 5$.

 The remainder of this paper is organized as follows. In Section~$2$ we introduce some preliminary results. In Section~$3$ and Section $4$, the weight distribution of $\mathcal{C}_f$ and the parameters of its dual are determined for $f(x)$ being a known PN-DO function and a Coulter-Matthews function, respectively. Section~$5$ concludes the paper.

\section{Preliminaries}

 Throughout this paper, we assume $m$ is a positive integer and adopt the following notation unless otherwise stated:
\begin{description}
\item{$\bullet$} $\bF_{p^m}$ is the finite field with $p^m$ elements and $\bF_{p^m}^*=\bF_{p^m}\setminus \{0\}$.
\item{$\bullet$} ${\rm Tr}(\cdot)$ is the absolute trace function from $\bF_{p^m}$ to $\bF_{p}$.
\item{$\bullet$} $\mathbf{0}_i$ is a vector with all entries being $0$, where $i$ is a nonnegative integer.
\item{$\bullet$} $\eta_0$ and  $\eta$ are the quadratic character over $\bF_p$ and $\bF_{p^m}$, respectively.
\end{description}

In the following, we recall some necessary preliminaries on quadratic forms over finite fields. A function~$f(x)$ from $\bF_{p^m}$ to $\bF_p$ can be viewed as an $m$-variable polynomial over $\bF_{p}$ if we identify the finite field $\bF_{p^m}$ with an $m$-dimensional vector space
$\bF_{p}^m$ over $\bF_{p}$.
The function $f(x)$ is called a quadratic form
if it is a homogenous polynomial of degree two as follows:
\[f(x_1, x_2, \cdots, x_m) = \sum_{1\leq i\leq j\leq m} a_{ij} x_ix_j , \,\, \, a_{ij}\in \bF_{p},\]
where we fix a basis of $\bF_p^m$ over $\bF_p$ and identify $x\in \bF_{p^m}$ with a vector $(x_1, x_2, \cdots, x_m)\in \bF_p^m$.
Then the rank of the quadratic
form $f(x)$ is defined as the codimension of $\bF_p$-vector space
\[ V = \{ x\in \bF_{p^m}\, \, |\,\, f(x+z)-f(x)-f(z) = 0, \,\, \mbox{for all}\,\, z\in \bF_{p^m} \} ,\]
which is denote by rank$(f)$. For a quadratic form $f(x)$ with $m$ variables over $\bF_p$, there exists a symmetric matrix $A$ such that $f(x)=XAX^T$,
where $X=(x_1, x_2, \cdots, x_m)\in \bF_p^m$ and $X^T$ denote the transpose of $X$. The determinant ${\rm det}(f)$ of $f(x)$
is defined to be the determinant of $A$, and $f(x)$ is non-degenerate if ${\rm det}(f)\neq 0$. There exists a
nonsingular matrix $M$ such that $MAM^T$ is a diagonal matrix. So, making a nonsingular linear substitution $X=YM$ with $Y=(y_1, y_2, \cdots, y_m)$ to the quadratic form $f(x)$, we have
 $$f(x)= YMAM^TY^T = \sum_{i=1}^r a_i y_i^2,\,\, a_i\in \bF_p, $$
where $r$ is the rank of $f(x)$. In \cite{Draperetal2007}, $\eta_0(\prod_{i=1}^r a_i)$ is called the type of $f(x)$. It is clear that $\eta_0({\rm det}(f))=\eta_0(\prod_{i=1}^r a_i)$ if  $f$ is a non-degenerate quadratic form. For more information about quadratic form, the readers can refer to~\cite{Lidl1983}.
The following lemma is a well known result about solutions of non-degenerate quadratic forms, which will be used to determine the weight distribution of $\mathcal{C}_f$.

\begin{lemma}\label{lem:quadraticsum1} (\cite[Theorems 6.26 and 6.27]{Lidl1983})
Let $f$ be a non-degenerate quadratic form in $m$ variables over $\bF_p$.
Define a function $\nu(\cdot)$ over $\bF_p$ by $\nu(0)=p-1$ and $\nu(\rho)=-1$ for $\rho \in \bF_p^*$.
Then for $b \in \bF_p$ the number of solutions of the equation $f(x_1, ...,x_m)=b$ in $\bF_p$ is
\begin{equation*}
p^{m-1}+\nu(b)p^{\frac{m-2}{2}}\eta_0((-1)^{\frac{m}{2}}\det(f))
\end{equation*}
for even $m$, and
\begin{equation*}
p^{m-1}+p^{\frac{m-1}{2}}\eta_0((-1)^{\frac{m-1}{2}}b\det(f))
\end{equation*}
for odd $m$, where $\eta_0$ is the quadratic character of $\bF_p$ and denote $\eta_0(0)=0$.
\end{lemma}

Let $\chi$ be the canonical additive character over $\bF_{p^m}$. The quadratic Gauss sum $G(\eta,\chi)$ is defined as
\[ G(\eta, \chi) = \sum_{x\in \mathbb{F}_{p^m}^*}\eta(x)\chi(x).\]
The possible values of the quadratic Gauss sums are given as follows.

\begin{lemma}\label{lem1}(\cite[Theorem 5.15]{Lidl1983})
Let $\mathbb{F}_{p^m}$ be a finite field, where $p$ is an odd prime and $m$ is a positive integer. Then
\[G(\eta,\chi)=\left\{ \begin{array}{lcl}
           (-1)^{m-1}p^{\frac{m}{2}}, & {\rm if}\, \,\, p\equiv 1 \pmod 4,\\
           (-1)^{m-1}\sqrt{-1}^m p^{\frac{m}{2}}, & {\rm if}\, \,\, p\equiv 3 \pmod 4. \end{array}  \right.\]
\end{lemma}


In order to obtain the parameters of the dual codes of the discussed subfield codes, we need the Pless power-moment
identities on linear codes. Let $\mathcal{C}$ be a $[n, k]$ code over $\mathbb{F}_p$, and denote its dual by $\mathcal{C}^{\perp}$. Let $A_i$ and $A^{\perp}_i$ denote the number of codewords with weight $i$ in $\mathcal{C}$ and $\mathcal{C}^{\perp}$, respectively. The first five Pless power-moment identities are as follows:
\begin{align*}
\sum_{i=0}^nA_i\,\,\,\,&=p^k;\\
\sum_{i=0}^niA_i\,\,&=p^{k-1}(pn-n-A_1^{\perp});\\
\sum_{i=0}^ni^{2}A_i&=p^{k-2}[(p-1)n(pn-n+1)-(2pn-p-2n+2)A_1^{\perp}+2A_2^{\perp}];\\
\sum_{i=0}^ni^{3}A_i&=p^{k-3}[(p-1)n(p^{2}n^{2}-2pn^{2}+3pn-p+n^{2}-3n+2)-(3p^{2}n^{2}-3p^{2}n-6pn^{2}+12pn\\
&+p^{2}-6p+3n^{2}-9n+6)A_1^{\perp}+(pn-p-n+2)A_2^{\perp}-6A_3^{\perp}];\\
\sum_{i=0}^ni^{4}A_i&=p^{k-4}[(p-1)n(p^{3}n^{3}-3p^{2}n^{3}+6p^{2}n^{2}-4p^{2}n+p^2+3pn^{3}-12pn^{2}+15pn-6p-n^{3}+6n^{2}-11n+6)\\
&-(4p^{3}n^{3}-6p^{3}n^{2}+4p^{3}n-p^{3}-12P^{2}n^{3}+36p^{2}n^{2}-38p^{2}n+14p^{2}+12pn^{3}-54pn^{2}+78pn-36p\\
&-4n^{3}+24n^{2}-44n+24)A_1^{\perp}+(12p^{2}n^{2}-24p^{2}n+14p^{2}-24pn^{2}+84pn-72p+12n^{2}-60n+72)A_2^{\perp}\\
&-(24pn-36p-24n+72)A_3^{\perp}+24A_4^{\perp}].
\end{align*}

 The following two lemmas on the bounds of linear codes are well-known.
\begin{lemma}(Sphere Packing bound)
Let $\mathcal{C}$ be a $p$-ary $[n, k,d]$ code. Then
$$p^n\geq p^k\sum_{i=0}^{\lfloor \frac{d-1}{2} \rfloor}\left(
\begin{array}{cccc}
   n  \\
     i  \\
\end{array}
\right)(p-1)^i.$$
\end{lemma}

\begin{lemma}\cite{Rouayheb2007}\label{bound2}
Let $q$ be an odd prime power and $A_q(n,d)$ be the maximum number of codewords of a $q$-ary code with length $n$ and Hamming distance at least $d$. If $q\geq 3$, $t=n-d+1$ and $r=\lfloor min\{\frac{n-t}{2}, \frac{t-1}{q-2}\}\rfloor$, then
\begin{equation*}
\begin{split}
A_q(n,d)\leq \frac{q^{t+2r}}{\sum_{i=0}^r\left(\begin{array}{cccc}
   t+2r  \\
     i  \\
\end{array}
\right)(q-1)^i}.
\end{split}
\end{equation*}
\end{lemma}


\section{The weight distribution of $\mathcal{C}_{f}$ for $f(x)$ being a  PN-DO function}

In this section, we determine the weight distribution of $\mathcal{C}_{f}$ for $f(x)$ being a PN-DO function from $\mathbb{F}_{p^m}$ to $\mathbb{F}_{p^m}$. Firstly, we show a
relationship between the determinants ${\rm det}({\rm Tr}(f(x)))$ and ${\rm det}({\rm Tr}(af(x)))$,  where $a \in \mathbb{F}_{p^m}^*$ and $f(x)$ is a DO polynomial over $\mathbb{F}_{p^m}$.
This result is general and interesting.

\begin{lemma}\label{lem:detrelation}
Let $a\in \mathbb{F}_{p^m}^{*}$ and $ f(x)$ be a  DO polynomial over $\mathbb{F}_{p^m}$, then  $${\rm det}({\rm Tr}(af(x)))=a^{\frac{p^m-1}{p-1}}{\rm det}({\rm Tr}(f(x))).$$
\end{lemma}
{\it Proof}.  Assume that $f(x)=\sum_{0\leq i,j\leq m-1}\delta_{ij}x^{p^i+p^j}$, where $\delta_{ij} \in \mathbb{F}_{p^m}$. Let $\{v_1, v_2, \cdots, v_m\}$ be a basis of $\bF_{p^m}$ over $\bF_p$
and each $x \in \bF_{p^m}$ be uniquely expressed as
$$ x=x_1v_1+ x_2v_2+\cdots + x_mv_m. $$
Then we obtain
$${\rm Tr}(af(x))=\sum_{1\le k,l \le m}\sum_{0\le  i,j \le m-1}{\rm Tr}(a\delta_{i,j}v_{k}^{p^i}v_{l}^{p^j})x_{k}x_{l}. $$
Let $A_{k}=\sum_{i=0}^{m-1}\beta_iv_k^{p^i}$ and $B_{l}=\sum_{j=0}^{m-1}\gamma_{j}v_{l}^{p^j}$ for $k, l\in \{ 1, 2, \cdots,m\}$ and $\delta_{ij}=\beta_{i}\gamma_{j}$ for some $\beta_i, \gamma_j \in \bF_{p^m}^*$, where
$i, j\in \{ 0, 1,\cdots, m-1\}$. According to the definition
of determinant of quadratic forms, we have
\begin{equation*}
\begin{split}
&{\rm det}({\rm Tr}(af(x)))\\
&={\rm det}\left\{ \left(\begin{array}{cccc}
    \sum\limits_{0\le  i,j \le m-1}{\rm Tr}\left(a\delta_{ij}v_{1}^{p^i}v_{1}^{p^j}\right) &    \sum\limits_{0\le  i,j \le m-1}{\rm Tr}\left(a\delta_{ij}v_{1}^{p^i}v_{2}^{p^j}\right)    & \cdots&\sum\limits_{0\le  i,j \le m-1}{\rm Tr}\left(a\delta_{ij}v_{1}^{p^i}v_{m}^{p^j}\right) \\
    \sum\limits_{0\le  i,j \le m-1}{\rm Tr}\left(a\delta_{ij}v_{2}^{p^i}v_{1}^{p^j}\right) &    \sum\limits_{0\le  i,j \le m-1}{\rm Tr}\left(a\delta_{ij}v_{2}^{p^i}v_{2}^{p^j}\right)   & \cdots&\sum\limits_{0\le  i,j \le m-1}{\rm Tr}\left(a\delta_{ij}v_{2}^{p^i}v_{m}^{p^j}\right)\\
    \vdots  &  \vdots  &  \ddots & \vdots \\
    \sum\limits_{0\le  i,j \le m-1}{\rm Tr}\left(a\delta_{ij}v_{m}^{p^i}v_{1}^{p^j}\right)&    \sum\limits_{0\le  i,j \le m-1}{\rm Tr}\left(a\delta_{ij}v_{m}^{p^i}v_{2}^{p^j}\right)   &\cdots&\sum\limits_{0\le  i,j \le m-1}{\rm Tr}\left(a\delta_{ij}v_{m}^{p^i}v_{m}^{p^j}\right)
\end{array}\right)  \right\} \\ \\
&={\rm det}\left\{ \left( \begin{array}{cccc}
     A_{1}&   A_{1}^{p}    &  \cdots & A_{1}^{p^{m-1}} \\
     A_{2}&   A_{2}^{p}    &  \cdots &A_{2}^{p^{m-1}} \\
     \vdots  &  \vdots  &  \ddots & \vdots \\
   A_{m}&   A_{m}^{p}    &  \cdots &A_{m}^{p^{m-1}} \\
\end{array}\right)
\left(\begin{array}{cccc}
     a&   0    & \cdots&0 \\
     0&   a^p    & \cdots&0 \\
    \vdots  &  \vdots  &  \ddots & \vdots \\
     0&   0   & \cdots&a^{p^{m-1}} \\
\end{array}\right)
\left(\begin{array}{cccc}
     B_{1}&   B_{2}    &  \cdots& B_{m} \\
     B_{1}^{p}&  B_{2}^{p}      &\cdots &B_{m}^{p} \\
     \vdots  &  \vdots  &  \ddots  &\vdots \\
     B_{1}^{p^{m-1}} &  B_{2}^{p^{m-1}}   &\cdots & B_{m}^{p^{m-1}} \\
\end{array}\right)\right\}\\ \\
&=a^{\frac{p^m-1}{p-1}} {\rm det} \left\{ \left(\begin{array}{cccc}
     A_{1}&   A_{1}^{p}    &  \cdots & A_{1}^{p^{m-1}} \\
     A_{2}&   A_{2}^{p}    &  \cdots & A_{2}^{p^{m-1}} \\
     \vdots  &  \vdots  &  \ddots & \vdots \\
   A_{m}&   A_{m}^{p}    &  \cdots & A_{m}^{p^{m-1}} \\
\end{array}\right)
\left(\begin{array}{cccc}
     B_{1}&   B_{2}    &  \cdots & B_{m} \\
     B_{1}^{p}&   B_{2}^{p}      &\cdots & B_{m}^{p} \\
     \vdots  &  \vdots  &  \ddots  &\vdots \\
     B_{1}^{p^{m-1}} &   B_{2}^{p^{m-1}}      &\cdots & B_{m}^{p^{m-1}} \\
\end{array}\right) \right\} \\ \\
&=a^{\frac{p^m-1}{p-1}}{\rm det}({\rm Tr}(f(x))).
\end{split}
\end{equation*}
This completes the proof.  $\square$

Let $Q(x)$ be a quadratic function from $\mathbb{F}_{p^m}$ to $\mathbb{F}_{p}$ and rank$(Q(x))=m$. Let
\begin{equation}\label{eq:ddser}
N_{b,c}=|\{x \in \mathbb{F}_{p^m}\,|\,Q(x)+{\rm Tr}(bx)=c\}|,
\end{equation}
then the possible values of $ N_{b,c}$ are given as follows.

\begin{lemma}\label{lemma1}
Let $N_{b,c}$ be defined in (\ref{eq:ddser}). When $(b,c)$ runs through $(\mathbb{F}_{p^m}, \mathbb{F}_p)$, then
\begin{equation*}
\begin{split}
N_{b,c}=\left\{ \begin{array}{llll}
           p^{m-1}, & {\rm occur}& p^{m} &{\rm times},\\
           p^{m-1}- p^{\frac{m-1}{2}},  & {\rm occur }&  \frac{1}{2}(p-1)p^{m}  &{\rm times},\\
           p^{m-1}+ p^{\frac{m-1}{2}},  & {\rm occur }&  \frac{1}{2}(p-1)p^{m}  &{\rm times}\\
 \end{array}  \right.
\end{split}
\end{equation*}
if $m$ is odd and
\begin{equation*}
\begin{split}
N_{b,c}=\left\{ \begin{array}{llll}
          p^{m-1}+\varepsilon_0(p-1)p^{\frac{m-2}{2}}, & {\rm occur}&p^m &{\rm times},\\
           p^{m-1}-\varepsilon_0 p^{\frac{m-2}{2}},  & {\rm occur }&(p-1)p^m & {\rm times} \end{array}  \right.
\end{split}
\end{equation*}
if $m$ is even, where $\varepsilon_0=(-1)^{\frac{m(p-1)}{4}}\eta_0({\rm det}(Q(x)))$.
\end{lemma}
{\it Proof}. Let $\{v_1, v_2, \cdots, v_m\}$ be a basis of $\bF_{p^m}$ over $\bF_p$.
Then $x \in \bF_{p^m}$ can be expressed as
$$ x=x_1v_1+ x_2v_2+\cdots + x_mv_m. $$
Making a nonsingular linear substitution to $Q(x)+{\rm Tr}(bx)=c$, we have
\begin{equation}\label{eq:diag}
\sum_{i=1}^ma_ix_i^2+\sum_{i=1}^mb_ix_i=c,
\end{equation}
where $a_i, b_i\in \bF_p$.  Note that the number of $x\in \bF_{p^m}$ satisfying $Q(x)+{\rm Tr}(bx)=0$
equals the number of the tuple $(x_1, x_2, \cdots, x_m)\in \bF_p^m$ satisfying (\ref{eq:diag}).
Let $x_i=z_i-\frac{b_i}{2a_i}$ for $1 \leq i \leq m$, then (\ref{eq:diag}) is equivalent to
 $$ \sum_{i=1}^ma_iz_i^2=\sum_{i=1}^m\frac{b_i^2}{4a_i}+c.$$
 The proof falls into the following two cases.

Case~1: $m$ is odd. In this case,
by Lemma~\ref{lem:quadraticsum1} we have
\begin{equation*}
N_{b,c}=p^{m-1}+p^{\frac{m-1}{2}}\eta_0\left(\sum_{i=1}^m\frac{b_i^2}{4a_i}+c\right)
\eta_0\left((-1)^{\frac{m-1}{2}}\prod_{i=1}^ma_i\right).
\end{equation*}
When $(b,c)$ runs through $( \mathbb{F}_{p^m}, \mathbb{F}_p)$, then
\begin{equation*}
\begin{split}
N_{b,c}=\left\{ \begin{array}{llll}
           p^{m-1}, & {\rm occur}& p^{m} &{\rm times},\\
           p^{m-1}- p^{\frac{m-1}{2}},  & {\rm occur }&  \frac{1}{2}(p-1)p^{m}  &{\rm times},\\
           p^{m-1}+ p^{\frac{m-1}{2}},  & {\rm occur }&  \frac{1}{2}(p-1)p^{m}  &{\rm times}.\\
 \end{array}  \right.
\end{split}
\end{equation*}

Case~2: $m$ is even. In this case,
by Lemma~\ref{lem:quadraticsum1} we have
\begin{equation*}
N_{b,c}=p^{m-1}+\nu\left(\sum_{i=1}^m\frac{b_i^2}{4a_i}+c\right)p^{\frac{m-2}{2}}
\eta_0\left((-1)^{\frac{m}{2}}\prod_{i=1}^ma_i\right),
\end{equation*}
where $\nu(\cdot)$ is defined in Lemma~\ref{lem:quadraticsum1}. It is easy to check that $\eta_0((-1)^{\frac{m}{2}})=(-1)^{\frac{m(p-1)}{4}}$ and $\eta_0({\rm det}(Q(x)))=\eta_0(\prod_{i=1}^ma_i)\neq 0$
since rank$(Q(x))=m$. When $(b,c)$ runs through $( \mathbb{F}_{p^m}, \mathbb{F}_p)$, we have
\begin{equation*}
\begin{split}
N_{b,c}=\left\{ \begin{array}{llll}
          p^{m-1}+\varepsilon_0(p-1)p^{\frac{m-2}{2}}, & {\rm occur}&p^m &{\rm times},\\
           p^{m-1}-\varepsilon_0 p^{\frac{m-2}{2}},  & {\rm occur }&(p-1)p^m & {\rm times}, \end{array}  \right.
\end{split}
\end{equation*}
where $\varepsilon_0=(-1)^{\frac{m(p-1)}{4}}\eta_0({\rm det}(Q(x)))$. $\square$

Recall that each codeword in $\mathcal{C}_{f}$ has the following form:
\begin{equation*}\label{codewordsss}
\mathbf{c}_f =\left(\left( {\rm Tr}(af(x)+bx)+c\right)_{x \in \mathbb{F}_{p^m}}, {\rm Tr}(a)\right),
\end{equation*}
where $a, b\in \bF_{p^m}$ and $c\in\bF_p$. In the following, we first determine the possible values of ${\rm wt_H}(\mathbf{c}_f)$ when $(a,b,c)$ runs through $(\mathbb{F}_{p^m},\mathbb{F}_{p^m},\mathbb{F}_{p})$.
\begin{lemma}\label{weight}
Let $f(x)$ be a PN-DO function from $\mathbb{F}_{p^m}$ to $\mathbb{F}_{p^m}$, $\mathbf{c}_f$ be a codeword in $\mathcal{C}_f$
and $N_{b,c}$ be defined in (\ref{eq:ddser}). Let $\sigma=1$ if ${\rm Tr}(a)\neq0$ and $\sigma=0$ if ${\rm Tr}(a)=0$.
Then the possible values of ${\rm wt_H}(\mathbf{c}_f)$ are
\begin{equation*}
\begin{split}
{\rm wt_H}(\mathbf{c}_f) &=  \left\{ \begin{array}{lll}
        0,          & {\rm if}\, \,\,c=0,a=0,b=0,\\
       p^m ,      & {\rm if}\, \,\,c\neq 0,a=0,b=0,\\
              (p-1)p^{m-1}+\sigma ,       &{\rm if}\, \,\, N_{b,c}=p^{m-1},\\
              (p-1)p^{m-1}-p^{\frac{m-1}{2}}+\sigma,  &{\rm if}\, \,\,  N_{b,c}=p^{m-1}+p^{\frac{m-1}{2}},\\
              (p-1)p^{m-1}+p^{\frac{m-1}{2}}+\sigma,  &{\rm if}\, \,\, N_{b,c}=p^{m-1}-p^{\frac{m-1}{2}}\end{array}  \right.\\\
\end{split}
\end{equation*}
 if $m$ is odd and
\begin{equation*}
\begin{split}
{\rm wt_H}(\mathbf{c}_f) &=  \left\{ \begin{array}{lll}
              0,          & {\rm if}\, \,\,c=0,a=0,b=0,\\
              p^m ,       & {\rm if}\, \,\,c\neq 0,a=0,b=0,\\
              (p-1)p^{m-1},  & {\rm if}\, \,\,a=0,b\neq 0,\\
              (p-1)(p^{m-1}-\varepsilon p^{\frac{m-2}{2}})+\sigma ,       &{\rm if}\, \,\, N_{b,c}=p^{m-1}+\varepsilon(p-1)p^{\frac{m-2}{2}} \,\,{\rm and}\,\, \eta(a)=1,\\
              (p-1)(p^{m-1}+\varepsilon p^{\frac{m-2}{2}})+\sigma ,       &{\rm if}\, \,\, N_{b,c}=p^{m-1}-\varepsilon(p-1)p^{\frac{m-2}{2}} \,\,{\rm and}\,\, \eta(a)=-1,\\
              (p-1)p^{m-1}+\varepsilon  p^{\frac{m-2}{2}}+\sigma,  &{\rm if}\, \,\,  N_{b,c}= p^{m-1}-\varepsilon p^{\frac{m-2}{2}}\,\,{\rm and}\,\, \eta(a)=1,\\
              (p-1)p^{m-1}-\varepsilon p^{\frac{m-2}{2}}+\sigma,  &{\rm if}\, \,\,  N_{b,c}= p^{m-1}+\varepsilon p^{\frac{m-2}{2}}\,\,{\rm and}\,\, \eta(a)=-1\end{array}  \right.\ \\
\end{split}
\end{equation*}
if $m$ is even, where $\varepsilon=(-1)^{\frac{m(p-1)}{4}}\eta_0({\rm det}({\rm Tr}(f(x))))$.
\end{lemma}
{\it Proof.}
When $a=0$, the Hamming weight of $\mathbf{c}_f$ can be easily determined and the result is given as follows:
\begin{equation*}
\begin{split}
{\rm wt_H}(\mathbf{c}_f) &=  \left\{ \begin{array}{lll}
              0,          & {\rm if}\, \,\,c=0,b=0,\\
              p^m ,       & {\rm if}\, \,\,c\neq 0,b=0,\\
              (p-1)p^{m-1},  & {\rm if}\, \,\,b\neq 0.\end{array}  \right.\ \\
\end{split}
\end{equation*}
 When $a\neq0$, from the expression of $\mathbf{c}_f$ we know
\begin{equation}\label{eq:sdas}
  {\rm wt_H}(\mathbf{c}_f)=p^m+\sigma-|\{x \in \mathbb{F}_{p^m}\,|\,{\rm Tr}\left(af(x)\right)+{\rm Tr}(bx)=-c\}|.
  \end{equation}
  By the definition of  PN-DO functions, it is easy to see that rank$({\rm Tr}(af(x)))=m$ for any $a \in \mathbb{F}_{p^m}^*$.
  If $m$ is odd, from Lemma \ref{lemma1} and (\ref{eq:sdas}) we have
\begin{equation*}
\begin{split}
{\rm wt_H}(\mathbf{c}_f) &=  \left\{ \begin{array}{lll}
              (p-1)p^{m-1}+\sigma ,       &{\rm if}\, \,\, N_{b,c}=p^{m-1},\\
              (p-1)p^{m-1}-p^{\frac{m-1}{2}}+\sigma,  &{\rm if}\, \,\,  N_{b,c}=p^{m-1}+p^{\frac{m-1}{2}},\\
              (p-1)p^{m-1}+p^{\frac{m-1}{2}}+\sigma,  &{\rm if}\, \,\, N_{b,c}=p^{m-1}-p^{\frac{m-1}{2}}.
              \end{array}  \right.\ \\
\end{split}
\end{equation*}
If $m$ is even, from Lemma~\ref{lem:detrelation} we have
\begin{equation*}
\begin{split}
\eta_0({\rm det}({\rm Tr}(af(x)))) &=  \left\{ \begin{array}{lll}
              \eta_0({\rm det}({\rm Tr}(f(x)))),       &{\rm if}\,\,a {\rm \,\, is}\,\, a {\rm \,\,square \,\,element \,\,in\,\, }\mathbb{F}_{p^m}^*,\\
               -\eta_0({\rm det}({\rm Tr}(f(x)))),       &{\rm if}\,\,a {\rm \,\, is}\,\, a {\rm \text{ non-square} \,\,element \,\,in\,\, }\mathbb{F}_{p^m}^*.
              \end{array}  \right.\ \\
\end{split}
\end{equation*}
Then from Lemma \ref{lemma1} and (\ref{eq:sdas}) we obtain
\begin{equation*}
\begin{split}
{\rm wt_H}(\mathbf{c}_f) &=  \left\{ \begin{array}{lll}
              (p-1)(p^{m-1}-\varepsilon p^{\frac{m-2}{2}})+\sigma ,       &{\rm if}\, \,\, N_{b,c}=p^{m-1}+\varepsilon(p-1)p^{\frac{m-2}{2}} \,\,{\rm and}\,\, \eta(a)=1,\\
              (p-1)(p^{m-1}+\varepsilon p^{\frac{m-2}{2}})+\sigma ,       &{\rm if}\, \,\, N_{b,c}=p^{m-1}-\varepsilon(p-1)p^{\frac{m-2}{2}} \,\,{\rm and}\,\, \eta(a)=-1,\\
              (p-1)p^{m-1}+\varepsilon  p^{\frac{m-2}{2}},  &{\rm if}\, \,\,  N_{b,c}= p^{m-1}-\varepsilon p^{\frac{m-2}{2}}\,\,{\rm and}\,\, \eta(a)=1,\\
              (p-1)p^{m-1}-\varepsilon  p^{\frac{m-2}{2}},  &{\rm if}\, \,\,  N_{b,c}= p^{m-1}+\varepsilon p^{\frac{m-2}{2}}\,\,{\rm and}\,\, \eta(a)=-1,
              \end{array}  \right.\ \\
\end{split}
\end{equation*}
where $\varepsilon=(-1)^{\frac{m(p-1)}{4}}\eta_0({\rm det}({\rm Tr}(f(x))))$. This completes the proof.  $\square$

To determine the frequency of each Hamming weight of $\mathcal{C}_f$, we still need the following result.

\begin{lemma}\cite[Lemma 14]{Hengar}\label{lemeven}
 Let $\mathbb{F}_{p^m}$ be the finite field of $p^m$ elements, where $p$ is an odd prime. Let $\eta$ be the quadratic multiplicative character of $\mathbb{F}_{p^m}$. If $m$ is even, then
\begin{equation*}
\begin{split}
\left|\left\{a \in \mathbb{F}_{p^m}^*\,|\, \eta(a)=\pm 1\,\, and\,\, {\rm Tr}(a)=0\right\}\right|=
\frac{p^{m-1}-1\mp (p-1)p^{\frac{m-2}{2}}(-1)^{\frac{(p-1)m}{4}}}{2},
\end{split}
\end{equation*}
and
\begin{equation*}
\begin{split}
\left|\{a \in \mathbb{F}_{p^m}^*\,|\, \eta(a)=\pm1\,\, and\,\, {\rm Tr}(a)\neq0\}\right|=
\frac{(p-1)(p^{m-1}\pm p^{\frac{m-2}{2}}(-1)^{{\frac{(p-1)m}{4}}})}{2}.
\end{split}
\end{equation*}
\end{lemma}

 With the above preparations, we now give the weight distribution of $\mathcal{C}_f$ and the parameters of its dual.
\begin{theorem}\label{Theorem 1}
Let $ f(x)$ be a PN-DO function from $\mathbb{F}_{p^m}$ to $\mathbb{F}_{p^m}$ and $\mathcal{C}_f$ be the linear code defined in (\ref{code0}). Then the following statements hold.
\begin{description}
\item{(1)} If $m$ is odd, then $\mathcal{C}_f$ is a $[p^m+1,2m+1,(p-1)p^{m-1}-p^{\frac{m-1}{2}}]$ code with weight distribution in Table \ref{Table1}. Its dual has parameters $[p^m+1,p^m-2m,4]$, which is distance-optimal with respect to the Sphere Packing bound for $p\geq 5$.
\begin{table}[h]
{\caption{\rm   The weight distribution of $\mathcal{C}_f$ for $m$ being odd}\label{Table1}
\begin{center}
\begin{tabular}{cccc}\hline
    $i$ & $A_i$ \\\hline
$0$  & $1$   \\
$(p-1)p^{m-1}+1$  & $(p^m-p^{m-1})p^m$   \\
$(p-1)p^{m-1}\pm p^{\frac{m-1}{2}}+1$  & $\frac{1}{2}(p^m-p^{m-1})(p-1)p^m$   \\
$p^m$  & $p-1$   \\
$(p-1)p^{m-1}$  & $p(p^m-1)+p^m(p^{m-1}-1)$   \\
$(p-1)p^{m-1}\pm p^{\frac{m-1}{2}}$  & $\frac{1}{2}(p^{m-1}-1)(p-1)p^m$   \\
   \hline
\end{tabular}
\end{center}}
\end{table}

\item{(2)} If $m$ is even, then $\mathcal{C}_f$ is a $[p^m+1,2m+1]$ code with weight distribution in Table \ref{Table2}, in which
$\varepsilon=(-1)^{m(p-1)/4}\eta_0({\rm det}({\rm Tr}(f(x))))$. Its dual has parameters $[p^m+1,p^m-2m,4]$, which is distance-optimal with
respect to the Sphere Packing bound for $p\geq 5$.
\begin{table}[h]
{\caption{\rm   The weight distribution of $\mathcal{C}_f $ for $m$ being even}\label{Table2}
\begin{center}
\begin{tabular}{cccc}\hline
 $i$ & $A_i$ \\\hline
 $0$  & $1$   \\
 $p^m$  & $p-1$   \\
 $(p-1)p^{m-1}$  & $p(p^m-1)$   \\
 $(p-1)(p^{m-1}\pm\varepsilon p^{\frac{m-2}{2}})+1$  & $\frac{1}{2}(p^m-p^{m-1}\mp(p-1)p^{\frac{m-2}{2}}(-1)^{\frac{1}{4}(p-1)m})p^m$   \\
 $(p-1)p^{m-1}\pm\varepsilon p^{\frac{m-2}{2}}+1$  & $\frac {1}{2}(p^{m}-p^{m-1}\pm (p-1)p^{\frac{m-2}{2}}(-1)^{\frac{1}{4}(p-1)m})(p-1)p^m$   \\
 $(p-1)(p^{m-1}\pm\varepsilon p^{\frac{m-2}{2}})$  & $\frac{1}{2}(p^{m-1}-1\pm(p-1)p^{\frac{m-2}{2}}(-1)^{\frac{1}{4}(p-1)m})p^m$   \\
 $(p-1)p^{m-1}\pm\varepsilon p^{\frac{m-2}{2}}$  & $\frac{1}{2}(p^{m-1}-1\mp(p-1)p^{\frac{m-2}{2}}(-1)^{\frac{1}{4}(p-1)m})(p-1)p^m$   \\
     \hline
\end{tabular}
\end{center}}
\end{table}
\end{description}
\end{theorem}
{\it Proof.}  Let $\mathbf{c}_f$  be a codeword in $\mathcal{C}_f$. It is easy to see that when $a=0$ and $(b,c)$ runs over $(\mathbb{F}_{p^m}, \mathbb{F}_p)$, the number of $\wt(\mathbf{c}_f)$ being $p^m$ or $(p-1)p^{m-1}$ is $p-1$ or $p(p^m-1)$, respectively. When $m$ is odd, the multiplicities of possible values of $\wt(\mathbf{c}_f)$ for $(a,b,c)$ running through $(\mathbb{F}_{p^m}^*,\mathbb{F}_{p^m}, \mathbb{F}_p)$ can be shown easily by Lemma \ref{lemma1}. In the following, we only prove the case for $m$ being even.

 Define
\begin{equation*}\label{eq:quadraticformT}
N_{\mu_0,\mu_1,\mu_2}=\left|\left\{(a,b,c) \in (\mathbb{F}_{p^m}^*,\mathbb{F}_{p^m},\mathbb{F}_p)\, |\, \wt(\mathbf{c}_f) = (p-1)p^{m-1}+\mu_0\varepsilon p^{\frac{m}{2}}+\mu_1\varepsilon p^{\frac{m-2}{2}}+\mu_2\right\}\right|,
\end{equation*}
where $\mu_0\in\{0,1,-1\}$, $\mu_1\in \{1,-1\}$ and $\mu_2 \in \{0,1\}$. By Lemmas \ref{lemma1}-\ref{lemeven} we have
\begin{equation*}
\begin{split}
N_{1,-1,0}&=\left|\left\{(a,b,c) \in (\mathbb{F}_{p^m}^*,\mathbb{F}_{p^m},\mathbb{F}_p)\, |\, \wt(\mathbf{c}_f) = (p-1)p^{m-1}+\varepsilon p^{\frac{m}{2}}-\varepsilon p^{\frac{m-2}{2}}\right\}\right|\\
&=\left|\left\{(a,b,c) \in (\mathbb{F}_{p^m}^*,\mathbb{F}_{p^m},\mathbb{F}_p)\,\,|\,{\rm Tr}(a)=0,\,\,\eta(a)=-1,\,\, N_{b,c}=p^{m-1}-\varepsilon(p-1)p^{\frac{m-2}{2}} \right\}\right|\\
&=\frac{1}{2}(p^{m-1}-1+(p-1)p^{\frac{m-2}{2}}(-1)^{\frac{1}{4}(p-1)m})p^m,
\end{split}
\end{equation*}
and
 \begin{equation*}
\begin{split}
N_{1,-1,1}&=\left|\left\{(a,b,c) \in (\mathbb{F}_{p^m}^*,\mathbb{F}_{p^m},\mathbb{F}_p)\, |\, \wt(\mathbf{c}(a,b,c)) = (p-1)p^{m-1}+\varepsilon p^{\frac{m}{2}}-\varepsilon p^{\frac{m-2}{2}}+1\right\}\right|\\
&=\left|\left\{(a,b,c) \in (\mathbb{F}_{p^m}^*,\mathbb{F}_{p^m},\mathbb{F}_p)\,\,|\,{\rm Tr}(a)\neq0,\,\,\eta(a)=-1,\,\, N_{b,c}=p^{m-1}-\varepsilon(p-1)p^{\frac{m-2}{2}} \right\}\right|\\
&=\frac{1}{2}(p^m-p^{m-1}-(p-1)p^{\frac{m-2}{2}}(-1)^{\frac{1}{4}(p-1)m})p^m.
\end{split}
\end{equation*}
 By similar calculations, we can get the values of $N_{-1,1,0}$, $N_{-1,1,1}$, $N_{0,-1,0}$, $N_{0,-1,1}$, $N_{0,1,0}$ and $N_{0,1,1}$. Then the  weight distribution of $\mathcal{C}_f$ is obtained in Table \ref{Table2}.

 From Table \ref{Table1}, Table \ref{Table2} and  the first five Pless power-moment identities, we know that the dual of $\mathcal{C}_f$ has parameters $[p^m+1,p^m-2m,4]$. By the Sphere Packing bound, we obtain
\begin{equation*}
\begin{split}
p^{p^m+1}\geq p^{p^m+1-(2m+1)}\left(\sum_{i=0}^{\lfloor\frac{{\rm d_H}(\mathcal{C}_f^{\perp})-1}{2}\rfloor}\left(
\begin{array}{cccc}
   p^m+1  \\
     i  \\
\end{array}
\right)(p-1)^i\right).
\end{split}
\end{equation*}
Then ${\rm d_H}(\mathcal{C}_f^{\perp})\leq 6$ if $p=3$ and ${\rm d_H}(\mathcal{C}_f^{\perp})\leq 4$ if $p\geq 5$. Hence, $\mathcal{C}_f^{\perp}$ is optimal with respect to the Sphere Packing bound if $p\geq 5$. $\square$

\begin{remark}
Assume that $f(x)=x^{p^k+1}$ is a PN function, where $k$ is a non-negative integer. By making a nonlinear substitution to $f(x)$
and using Lemma~\ref{lem1}, we have
\begin{equation*}
\begin{split}
\sum_{x\in \bF_{p^m}}\zeta_p^{{\rm Tr}(f(x))}= \sum_{x_1,\cdots,x_m \in \bF_p}\zeta_p^{a_1 x_1^2+\cdots +a_m x_m^2}= \eta_0(\prod_{i=1}^m a_i) (\sqrt{-1})^{\frac{m}{4}(p-1)^2}p^{\frac{m}{2}},
\end{split}
\end{equation*}
where $a_i\in \bF_p$ for $i=1,2,\cdots,m$. From \cite[Theorem 1]{Coulter19980}, we see $\eta_0({\rm det}({\rm Tr}(f(x))))=-1$ since $\eta_0({\rm det}({\rm Tr}(f(x))))=\eta_0(\prod_{i=1}^m a_i)$. Substituting the value of $\varepsilon=-(-1)^{\frac{m(p-1)}{4}}$ into Table \ref{Table2}, then we obtain \cite[Theorem 16]{Hengar} and \cite[Theorem 12]{Wang2020}. This means
that those results can be seen as a special case of Theorem \ref{Theorem 1}.
\end{remark}

\begin{example}\label{example1}
Let $\mathcal{C}_f$ be the linear code in Theorem \ref{Theorem 1}.
\begin{description}
\item{(1)} If $p=5$, $m=2$, $f(x)=x^2$ or $x^2-x^{10}+x^6$, then $\eta_0({\rm det}({\rm Tr}(f(x))))=-1$. So, $\mathcal{C}_f $ has parameters $[26,5,16]$ and weight enumerator $1+100x^{16}+200x^{17}+1320x^{20}+400x^{21}+800x^{22}+304x^{25}.$
Its dual has parameters $[26,21,4]$.
\item{(2)} If $p=5$, $m=2$, $f(x)=\xi x^2$ or $\xi(x^2-x^{10}+x^6)$ for $\xi$ being a primitive element in $\mathbb{F}_{5^2}$, then $\eta_0({\rm det}({\rm Tr}(f(x))))=1$. So, $\mathcal{C}_f $ has parameters $[26,5,17]$ and weight enumerator $1+300x^{17}+400x^{19}+920x^{20}+1200x^{22}+24x^{100}+204x^{25}.$
Its dual has parameters $[26,21,4]$.
 \item{(3)} If $p=3$, $m=3$, $f(x)=x^4$ or $x^{10}-x^6-x^2$,  then $\mathcal{C}_f$ has parameters $[28,7,15]$ and weight enumerator $1+216x^{15}+486x^{16}+294x^{18}+486x^{19}+216x^{21}+486x^{22}+2x^{27}$.
 Its dual has parameters $[28,21,4]$.
\end{description}
All of these codes and their duals are optimal or almost optimal with respect to the code tables in \cite{MGrassl}. These results have been verified by Magma.
\end{example}

By deleting the last coordinate of the codewords of $\mathcal{C}_f$, the punctured code of $\mathcal{C}_f$ has the form
\begin{equation}\label{r34wrwsf}
\begin{split}
\bar{\mathcal{C}}_f=\left\{\left(\left({\rm Tr}\left(af(x)+bx\right)+c\right)_{x \in \mathbb{F}_{p^m}}\right) : \,\, a, b \in \mathbb{F}_{p^m}, c \in \mathbb{F}_p\right\},
\end{split}
\end{equation}
where $f(x)$ is a PN-DO function from $\mathbb{F}_{p^m}$ to $\mathbb{F}_{p^m}$.  The weight distribution of this punctured code can be directly derived
from Table~\ref{Table1} and Table~\ref{Table2}, which is the same with the results presented in \cite[Theorem 2]{Li2009}.
\begin{corollary}\label{rwdc}
Let $ f(x)$ be a PN-DO function from $\mathbb{F}_{p^m}$ to $\mathbb{F}_{p^m}$ and $\bar{\mathcal{C}}_f$ be the linear code defined in (\ref{r34wrwsf}). Let $\bar{\mathcal{C}}^{\perp}_f$ denote the dual of $\bar{\mathcal{C}}_f$, then the following statements hold.
\begin{description}
\item{(1)} If $m$ is odd,  then $\bar{\mathcal{C}}_f$ is a $[p^m,2m+1,(p-1) p^{m-1}-p^{\frac{m-1}{2}}]$ code with the weight distribution as follows:
    \begin{equation*}
\begin{split}
\left\{ \begin{array}{llll}
               0,   & \text{ occur} & 1 & \text{ time,}\\
            (p-1)p^{m-1},  & \text{ occur} & (p^{m}-1)(p^{m-1}+1)p & \text{ times,}\\
              (p-1)p^{m-1}\pm p^{\frac{m-1}{2}},  & \text{ occur} &  \frac{1}{2}(p^{m}-1)p^m(p-1) & \text{ times,}\\
            p^m, & \text{ occur}   & p-1 & \text{ times.}\\
              \end{array}  \right.\ \\
\end{split}
\end{equation*}

\item{(2)} If $m$ is even,  then $\bar{\mathcal{C}}_f$ is a $[p^m,2m+1,(p-1)(p^{m-1}-p^{\frac{m-2}{2}})]$ code with the weight distribution as follows:
\begin{equation*}
\begin{split}
\left\{ \begin{array}{llll}
               0,   & \text{ occur} & 1 & \text{ time,}\\
           (p-1)p^{m-1},  & \text{ occur} & (p^{m}-1)p & \text{ times,}\\
              (p-1)p^{m-1}\pm (p-1)p^{\frac{m-2}{2}},  & \text{ occur} & \frac{1}{2}(p^{m}-1)p^{m} & \text{ times,}\\
             (p-1)p^{m-1}\pm p^{\frac{m-2}{2}},  & \text{ occur} &  \frac{1}{2}(p^{m}-1)(p-1)p^{m} & \text{ times,}\\
            p^m, & \text{ occur}   & p-1 & \text{ times.}\\
              \end{array}  \right.\ \\
\end{split}
\end{equation*}
\item{(3)} If $p=3$ and $m>1$, $\bar{\mathcal{C}}^{\perp}_f$ has parameters $[3^m,3^m-2m-1,5]$, which is distance-optimal with respect to Lemma~\ref{bound2}. If $p>3$, $\bar{\mathcal{C}}^{\perp}_f$ has parameters $[p^m,p^m-2m-1,4]$, which is distance-optimal with respect to the Sphere Packing bound.
\end{description}
\end{corollary}
{\it Proof}. The weight distribution of $\bar{\mathcal{C}}_f$ can be derived easily from Theorem \ref{Theorem 1} and we omit the details
here. By the first five Pless power-moment identities, we obtain ${\rm d_H}(\bar{\mathcal{C}}^{\perp}_f)> 4$ if $p=3$ and ${\rm d_H}(\bar{\mathcal{C}}^{\perp}_f)= 4$ if $p\geq 5$.
From the Sphere Packing bound, we know that $\bar{\mathcal{C}}^{\perp}_f$ is distance-optimal with respect to the Sphere Packing bound if $p\geq 5$ and ${\rm d_H}(\bar{\mathcal{C}}^{\perp}_f)\leq6$ if $p=3$ and $m>1$ (If $m=1$ and $p=3$, then $\mathcal{\bar{C}}_f$ is a $[3,3,1]$ code and $\mathcal{\bar{C}}^{\perp}_f=\mathbf{0}$).

Assume that there exists a ternary linear code with parameters $[3^m, 3^m-2m-1,6]$. Applying Lemma \ref{bound2}, we have $q=3$, $n=3^m$, $t=3^m-5$, $r=2$, and
    $$3^{3^m-2m-1}\leq \frac{3^{3^m-1}}{1+2(3^m-1)^2},$$
    which is impossible since $m>1$. Hence, ${\rm d_H}(\bar{\mathcal{C}}^{\perp}_f)=5$ and $\bar{\mathcal{C}}^{\perp}_f$ is optimal with respect to Lemma \ref{bound2} if $p=3$. $\square$

\begin{example}\label{example2}
Let $\bar{\mathcal{C}}_f$ be the linear code in Corollary \ref{rwdc}.
\begin{description}
\item{(1)} If $p=5,$ $m=2$, $f(x)=x^2$ or $x^2-x^{10}+x^6$, then $\bar{\mathcal{C}_f}$ has parameters $[25,5,16]$ and weight enumerator $1+300x^{16}+1200x^{19}+120x^{20}+1200x^{21}+300x^{24}+4x^{25}.$
Its dual has parameters $[26,21,4]$.
\item{(3)} If $p=3,$ $m=3$, $f(x)=x^2$ or $x^{10}-x^{6}-x^2$, then $\bar{\mathcal{C}_f}$ has parameters $[27,7,15]$ and weight enumerator $1+702x^{15}+780x^{18}+702x^{21}+2x^{27}$.
 Its dual has parameters $[27,20,5]$.
\end{description}
All of these codes and their duals are optimal or almost optimal with respect to the code tables in \cite{MGrassl}.  These results have been verified by Magma.
\end{example}

\section{The weight distribution of $\mathcal{C}_{f}$ for $f(x)$ being a Coulter-Matthews function}

In this section, we determine the weight distribution of $\mathcal{C}_{f}$ for $f(x)$ being a Coulter-Matthews function from $\mathbb{F}_{3^m}$ to $\mathbb{F}_{3^m}$ and investigate the parameters of the dual of $\mathcal{C}_f$, where $f(x)=x^{\frac{3^{k}+1}{2}}$ for $ 2\nmid k$ and gcd$(m,k)=1$. We start with the following lemma, which is the key for us to determine the number of codewords with weight $4$ in the dual of ${\mathcal{C}}_f$.
\begin{lemma}\label{eq:dddfg}
Let $f(x)=x^{\frac{3^{k}+1}{2}}$, where $ 2\nmid k$ and gcd$(m,k)=1$. Let $u \in \mathbb{F}_{3^m}$ and $N_u$ denote the number of solutions of the system of equations:
\begin{equation}\label{eq:objeq}
\begin{split}
 \left\{ \begin{array}{lll}
              x+y+z=0,\\
              f(x)+f(y)+f(z)+u=0.
              \end{array}  \right.\ \\
\end{split}
\end{equation}
When $(x,y,z)$ runs over $\mathbb{F}_{3^m}^3$, then
\begin{equation*}
\begin{split}
 N_u=\left\{ \begin{array}{lll}
               3^m,   & {\text if\,\, u=0,}\\
             2\cdot3^m,   & {\rm if }\,\, u \,\, {\rm is \,\,a \,\,square \,\,element \,\,in}\,\, \mathbb{F}_{3^m}, \\
             0,   & {\rm  if}\,\, u \,\, {\rm is} \,\, {\rm a \,\, nonsqaure \,\,element \,\,in\,\,} \mathbb{F}_{3^m}. \\
              \end{array}  \right.\ \\
\end{split}
\end{equation*}
\end{lemma}
{\it Proof}. If $u=0$, the result is given by \cite[Lemma~5]{Ness2006} and $N_0=3^m$. In the following, we only prove the result for $u\neq 0$ and the proof falls into two cases.

Case 1: $u$ is a non-square element in  $\mathbb{F}_{3^m}$.  If $z=0$, then we have $x^{\frac{3^k+1}{2}}=u$, which is impossible
since gcd$(\frac{3^k+1}{2}, 3^m-1)=2$. Hence, $N_u=0$ for $z=0$. If $z\neq 0$, when  $(x,y,z)$ runs over $\mathbb{F}_{3^m}^3$, the value of $N_u$ is equal to the number of solutions of the system of equations:
\begin{equation}\label{eq:xy1}
\begin{split}
 \left\{ \begin{array}{lll}
              x+y+1=0,\\
              z^{\frac{3^k+1}{2}}(x^{\frac{3^k+1}{2}}+y^{\frac{3^k+1}{2}}+1)+u=0.
              \end{array}  \right.\ \\
\end{split}
\end{equation}
Substituting the first equation into the second equation of (\ref{eq:xy1}), we obtain
$$z^{\frac{3^k+1}{2}}(x^{\frac{3^k+1}{2}}+(-x-1)^{\frac{3^k+1}{2}}+1)=-u.$$
Replacing $x$ by $x+1$, we have
\begin{equation}\label{eqdasd}
z^{\frac{3^k+1}{2}}((x+1)^{\frac{3^k+1}{2}}+(x-1)^{\frac{3^k+1}{2}}+1)=-u.
\end{equation}
Let $\theta$ be a root of $t^2-xt+1=0$, then $\theta \in \mathbb{F}_{3^{2m}}$ and $x=\theta+\theta^{-1}$. Hence, the equation (\ref{eqdasd}) is reduced to
\begin{equation}\label{eq:theta}
z^{\frac{3^k+1}{2}}((\theta+\theta^{-1}+1)^{\frac{3^k+1}{2}}+(\theta+\theta^{-1}-1)^{\frac{3^k+1}{2}}+1)=-u.
\end{equation}
Multiplying $\theta^{\frac{3^k+1}{2}}$ on both sides of (\ref{eq:theta}), we have
$$z^{\frac{3^k+1}{2}}((\theta^2-2\theta+1)^{\frac{3^k+1}{2}}+(\theta^2+2\theta+1)^{\frac{3^k+1}{2}}+\theta^{\frac{3^k+1}{2}})=-u\theta^{\frac{3^k+1}{2}},$$
which implies
$$(\theta^{-1}z)^{\frac{3^k+1}{2}}(2(\theta^{3^k+1}+1)+\theta^{\frac{3^k+1}{2}})=-u,$$
and then
\begin{equation}\label{eq:thetasq}
(\theta^{-1}z)^{\frac{3^k+1}{2}}(\theta^{\frac{3^k+1}{2}}+1)^2=z^{\frac{3^k+1}{2}}(\theta^{\frac{3^k+1}{4}}+\theta^{-\frac{3^k+1}{4}})^2=u.
\end{equation}
Note that
$$ \theta^{\frac{3^k+1}{4}}+\theta^{-\frac{3^k+1}{4}}= D_{\frac{3^k+1}{4}}\left( \theta+\theta^{-1}, 1\right),$$
where $D_n(x, 1)\in \bF_{3^m}[x]$ is a Dickson polynomial of the first kind \cite[Chapter 7]{Lidl1983}.  Then $\theta^{\frac{3^k+1}{4}}+\theta^{-\frac{3^k+1}{4}}\in \mathbb{F}_{3^m}$ since $\theta+\theta^{-1}\in \mathbb{F}_{3^m}$. This means that $z^{\frac{3^k+1}{2}}(\theta^{\frac{3^k+1}{4}}+\theta^{-\frac{3^k+1}{4}})^2$ is a square element in $\mathbb{F}_{3^m}$ since $\frac{3^k+1}{2}$
is even. But $u$ is a non-square element in $\mathbb{F}_{3^m}$. So, there doesn't exist $\theta$ such that (\ref{eq:thetasq}) holds, i.e., there don't exist $x$, $y$, $z$ such that $(\ref{eq:objeq})$ holds.

Case 2: $u$ is a square element in $\mathbb{F}_{3^m}$. In this case, it is easy to show that $N_u$ is equal to the number of solutions of the system of equations:
\begin{equation}
\begin{split}
 \left\{ \begin{array}{lll}
              x+y+z=0,\\
              x^{\frac{3^k+1}{2}}+y^{\frac{3^k+1}{2}}+z^{\frac{3^k+1}{2}}+t^{\frac{3^k+1}{2}}u=0,
              \end{array}  \right.\ \\
\end{split}
\end{equation}
where $t \in \mathbb{F}_{3^m}^*$. When $t$ runs over $\mathbb{F}_{3^m}^*$, $ut^{\frac{3^k+1}{2}}$ runs over all square elements in $\bF_{3^m}^*$ since gcd$(\frac{3^k+1}{2}, 3^m-1)=2$.
So, $N_u$ is the same for all square elements $u$ in $\bF_{3^m}^*$. From Case~1 we have
\begin{equation*}
\begin{split}
3^{2m}=|\left\{ (x,y,z) \in \mathbb{F}_{3^{3m}}\,|\, x+y+z=0 \right\}|=\sum_{u \in {\rm SQ} }N_u+N_0= \frac{3^m-1}{2} N_u + 3^m,
\end{split}
\end{equation*}
where SQ is the set of the square elements in $\mathbb{F}_{3^m}^*$. Hence, $N_u=2\cdot 3^{m}$. $\square$

The following two lemmas will determine the number of codewords with weight $i$ in the dual of ${\mathcal{C}}_f$ for $1\leq i\leq 4$.
\begin{lemma}\label{lem:a1230}
Let $\mathcal{C}_f$ be the linear code defined in (\ref{code0}) and $\mathcal{C}_f^{\perp}$ denote its dual over $\mathbb{F}_{3}$. Let $f(x)=x^{\frac{3^{k}+1}{2}}$, where $ 2\nmid k$ and $\gcd(m,k)=1$. Assume that $A^{\perp}_i$ is the number of codewords with weight $i$ in $\mathcal{C}_f^{\perp}$. Then $A^{\perp}_1=A^{\perp}_2=A^{\perp}_3=0$.
\end{lemma}
 {\it Proof.}  We prove this result from the following three cases.

 Case 1: Assume that there is a codeword $\mathbf{c}^{\perp} \in \mathcal{C}_f^{\perp}$ with ${\rm wt_H}(\mathbf{c}^{\perp})=1$. Then $\mathbf{c}^{\perp}$ must have the form $(\mathbf{0},\alpha)$ since $(\mathbf{1},0) \in \mathcal{C}_f$, where $\alpha \in \mathbb{F}_3^*$, $\mathbf{0}$ and $\mathbf{1}$ are vectors of length $3^m$ with all entries being $0$ and $1$, respectively.
 However, for any ${\rm Tr}(a)\neq 0$, there is a codeword  $(\mathbf{u},{\rm Tr}(a)) \in \mathcal{C}_f$ such that $(\mathbf{u},{\rm Tr}(a))(\mathbf{0},\alpha)\neq 0$, where $\mathbf{u}$ is a vector of length $3^m$. Hence, $A^{\perp}_1=0$.

Case 2: Assume that there is a codeword $\mathbf{c}^{\perp} \in \mathcal{C}_f^{\perp}$ with ${\rm wt_H}(\mathbf{c}^{\perp})=2$. Then $\mathbf{c}^{\perp}$ must have the form $(\mathbf{0}_1,\alpha,\mathbf{0}_2,-\alpha,\mathbf{0}_3)$ or  $(\alpha,\mathbf{0}_1,-\alpha,\mathbf{0}_2)$ since $(\mathbf{1},0) \in \mathcal{C}_f$, where $\alpha \in \mathbb{F}_3^*$. Recall that a codeword in $\mathcal{C}_f$ has the form
\begin{equation}\label{eq:codeform}
 \mathbf{c}_f =\left(\left( {\rm Tr}(af(x)+bx)+c\right)_{x \in \mathbb{F}_{p^m}}, {\rm Tr}(a)\right).
\end{equation}
Set $c=0$.  Then there exist two distinct elements $x, y\in \mathbb{F}_{3^m}$ such that
$$\alpha {\rm Tr}(ax^{\frac{3^k+1}{2}}+bx)-\alpha {\rm Tr}(a y^{\frac{3^k+1}{2}}+by)=0,$$
i.e.,
\[{\rm Tr}\left(a(x^{\frac{3^k+1}{2}}+y^{\frac{3^k+1}{2}})+b(x+y)\right)=0\]
for any $a, b \in \mathbb{F}_{p^m}.$ This is impossible. Hence, $A^{\perp}_2=0$.

Case 3: Assume that there is a codeword $\mathbf{c}^{\perp} \in \mathcal{C}_f^{\perp}$ with ${\rm wt_H}(\mathbf{c}^{\perp})=3$.
If the last entry of $\mathbf{c}^{\perp}$ is zero, then $\mathbf{c}^{\perp}$ must have the form $(\mathbf{0}_1,\alpha_1,\mathbf{0}_2,\alpha_2,\mathbf{0}_3,\alpha_3,\mathbf{0}_4)$ or $(\alpha_1,\mathbf{0}_1,\alpha_2,\mathbf{0}_2,\alpha_3,\mathbf{0}_3)$, where $\alpha_1,\alpha_2,\alpha_3 \in \mathbb{F}_{3}^*$. Since a codeword $\mathbf{c}_f$ in $\mathcal{C}_f$ has the form (\ref{eq:codeform}) and $(\mathbf{1},0) \in \mathcal{C}_f$,
there exist three distinct elements $x, y, z \in \mathbb{F}_{3^m}$ such that
\begin{equation*}
\begin{split}
 \left\{ \begin{array}{lll}
             \alpha_1x^{\frac{3^k+1}{2}}+\alpha_2y^{\frac{3^k+1}{2}}+\alpha_3z^{\frac{3^k+1}{2}}=0,\\
              \alpha_1x+\alpha_2y+\alpha_3z=0,\\
            \alpha_1+\alpha_2+\alpha_3=0,
 \end{array}  \right.\ \\
\end{split}
\end{equation*}
i.e.,
\begin{equation}\label{eq:dddf}
\begin{split}
 \left\{ \begin{array}{lll}
              x^{\frac{3^k+1}{2}}+y^{\frac{3^k+1}{2}}+z^{\frac{3^k+1}{2}}=0,\\
x+y+z=0.
              \end{array}  \right.\ \\
\end{split}
\end{equation}
From Lemma \ref{eq:dddfg} we know that  (\ref{eq:dddf}) holds if and only if $x=y=z$. Hence, there doesn't exist $\mathbf{c}^{\perp} \in \mathcal{C}_f^{\perp}$ with ${\rm wt_H}(\mathbf{c}^{\perp})=3$ if the last entry of $\mathbf{c}^{\perp}$ is zero.

If the last entry of $\mathbf{c}^{\perp}$ is nonzero, then $\mathbf{c}^{\perp}$ must have the form  $(\mathbf{0}_1,\alpha_1,\mathbf{0}_2,\alpha_2,\mathbf{0}_3,\alpha_3)$ or $(\alpha_1,\mathbf{0}_1,\alpha_2,\mathbf{0}_2,\alpha_3)$. So, there exist two distinct elements $x,y\in \mathbb{F}_{3^m}$ such that
\begin{equation*}
\begin{split}
 \left\{ \begin{array}{lll}
              \alpha_1x^{\frac{3^k+1}{2}}+\alpha_2y^{\frac{3^k+1}{2}}+\alpha_3=0,\\
\alpha_1x+\alpha_2y=0,\\
              \alpha_1+\alpha_2=0,
              \end{array}  \right.\ \\
\end{split}
\end{equation*}
which is impossible. Hence, $A^{\perp}_3=0$.   $\square$


\begin{lemma}\label{lem:a4}
Follow the notation introduced above. Then $A^{\perp}_4=4\cdot3^{m-1}$ if $m$ is even and $A^{\perp}_4=2\cdot3^{m-1}$  if $m$ is odd.
\end{lemma}
 {\it Proof.}
Assume that there is a codeword $\mathbf{c}^{\perp} \in \mathcal{C}_f^{\perp}$ with ${\rm wt_H}(\mathbf{c}^{\perp})=4$.
If the last entry of $\mathbf{c}^{\perp}$ is zero, then $\mathbf{c}^{\perp}$ must have the form $(\mathbf{0}_1,\alpha_1,\mathbf{0}_2,\alpha_2,\mathbf{0}_3,\alpha_3,\mathbf{0}_4, \alpha_4, \mathbf{0}_5)$ or $(\alpha_1,\mathbf{0}_1,\alpha_2,\mathbf{0}_2,\alpha_3,\mathbf{0}_3,\alpha_4,\mathbf{0}_4)$, where $\alpha_1,\alpha_2,\alpha_3,\alpha_4 \in \mathbb{F}_{3}^*$.
Since a codeword $\mathbf{c}_f$ in $\mathcal{C}_f$ has the form (\ref{eq:codeform}) and $(\mathbf{1},0) \in \mathcal{C}_f$, there exist four distinct elements $x,y,z,u\in \mathbb{F}_{3^m}$ such that
\begin{equation}\label{eq:rewf}
\begin{split}
 \left\{ \begin{array}{lll}
             \alpha_1x^{\frac{3^k+1}{2}}+\alpha_2y^{\frac{3^k+1}{2}}+\alpha_3z^{\frac{3^k+1}{2}}+\alpha_4u^{\frac{3^k+1}{2}}=0,\\
              \alpha_1x+\alpha_2y+\alpha_3z+\alpha_4u=0,\\
            \alpha_1+\alpha_2+\alpha_3+\alpha_4=0.
              \end{array}  \right.\ \\
\end{split}
\end{equation}
Since $\alpha_1, \alpha_2, \alpha_3, \alpha_4\in \bF_3^*$, from the last equation of (\ref{eq:rewf}), we know that two of $\alpha_1,\alpha_2,\alpha_3,\alpha_4$ must be the same. Without loss of generality, we assume $\alpha_1=\alpha_2$ and $\alpha_3=\alpha_4$. Then (\ref{eq:rewf}) is reduced to
\begin{equation}\label{eq:pnf}
\begin{split}
 \left\{ \begin{array}{lll}
x^{\frac{3^k+1}{2}}-z^{\frac{3^k+1}{2}}=u^{\frac{3^k+1}{2}}-y^{\frac{3^k+1}{2}},\\
x-z=y-u.
              \end{array}  \right.\ \\
\end{split}
\end{equation}
Let $x-z=\beta$. Then (\ref{eq:pnf}) is reduced to
\begin{equation*}
\begin{split}
(z+\beta)^{\frac{3^k+1}{2}}-z^{\frac{3^k+1}{2}}=(y+\beta)^{\frac{3^k+1}{2}}-y^{\frac{3^k+1}{2}}.
\end{split}
\end{equation*}
This is impossible since $x^{\frac{3^k+1}{2}}$ is a PN function and $y\neq z$.  Hence, there doesn't exist $\mathbf{c}^{\perp} \in \mathcal{C}_f^{\perp}$ with ${\rm wt_H}(\mathbf{c}^{\perp})=4$ if the last entry of $\mathbf{c}^{\perp}$ is zero.

If the last entry of $\mathbf{c}^{\perp}$ is nonzero, then $\mathbf{c}^{\perp}$ has the form  $(\mathbf{0}_1,\alpha_1,\mathbf{0}_2,\alpha_2,\mathbf{0}_3,\alpha_3,\mathbf{0}_4,\alpha_4)$ or $(\alpha_1,\mathbf{0}_1,\alpha_2,\mathbf{0}_2,\alpha_3,\mathbf{0}_3,\alpha_4)$. There exist three distinct elements $x,y,z \in \mathbb{F}_{3^m}$ such that
\begin{equation}\label{eq:rew1f}
\begin{split}
 \left\{ \begin{array}{lll}
              \alpha_1x^{\frac{3^k+1}{2}}+\alpha_2y^{\frac{3^k+1}{2}}+\alpha_3z^{\frac{3^k+1}{2}}+\alpha_4=0,\\
                            \alpha_1x+\alpha_2y+\alpha_3z=0,\\
              \alpha_1+\alpha_2+\alpha_3=0.
              \end{array}  \right.\ \\
\end{split}
\end{equation}
From the last equation of (\ref{eq:rew1f}), we know that $\alpha_1,\alpha_2,\alpha_3$ must be the same. So, the possible values of $(\alpha_1,\alpha_2,\alpha_3,\alpha_4)$ are $(1,1,1,1)$, $(1,1,1,2)$, $(2,2,2,1)$ and $(2,2,2,2)$. If $(\alpha_1,\alpha_2,\alpha_3,\alpha_4)=(1,1,1,1)$ or $(\alpha_1,\alpha_2,\alpha_3,\alpha_4)=(2,2,2,2)$, then (\ref{eq:rew1f}) becomes
\begin{equation}\label{eq:T}
\begin{split}
 \left\{ \begin{array}{lll}
             x^{\frac{3^k+1}{2}}+y^{\frac{3^k+1}{2}}+z^{\frac{3^k+1}{2}}+1=0,\\
              x+y+z=0.
              \end{array}  \right.\ \\
\end{split}
\end{equation}
From Lemma \ref{eq:dddfg}, the number of $(x,y,z) \in \mathbb{F}_{3^m}^3$ satisfying (\ref{eq:T}) is $2\cdot 3^m$.  If $(\alpha_1,\alpha_2,\alpha_3,\alpha_4)=(1,1,1,2)$ or $(\alpha_1,\alpha_2,\alpha_3,\alpha_4)=(2,2,2,1)$, then (\ref{eq:rew1f}) becomes
\begin{equation}\label{eq:1T}
\begin{split}
 \left\{ \begin{array}{lll}
             x^{\frac{3^k+1}{2}}+y^{\frac{3^k+1}{2}}+z^{\frac{3^k+1}{2}}+2=0,\\
              x+y+z=0.
              \end{array}  \right.\ \\
\end{split}
\end{equation}
It is clear that $2$ is square or non-square element in $\mathbb{F}_{3^m}$ if $m$ is even or odd, respectively. From Lemma \ref{eq:dddfg}, the number of $(x,y,z) \in \mathbb{F}_{3^m}^3$ satisfying (\ref{eq:1T}) is $2\cdot 3^m$ or $0$ if $m$ is even or odd, respectively.

Hence, the number of $(x,y,z) \in \mathbb{F}_{3^m}^3$ satisfying (\ref{eq:rew1f}) is $8\cdot 3^m$ if $m$ is even and $4\cdot 3^m$ if $m$ is odd, respectively. This means that $A^{\perp}_4=4\cdot3^{m-1}$ if $m$ is even and $A^{\perp}_4=2\cdot3^{m-1}$  if $m$ is odd.   $\square$

By deleting the last coordinate of the codewords of $\mathcal{C}_f$, we get the punctured code of $\mathcal{C}_f$ as follows:
\begin{equation}\label{rwrwsf}
\begin{split}
\bar{\mathcal{C}}_f=\left\{\left(\left({\rm Tr}(ax^{\frac{3^k+1}{2}}+bx)+c\right)_{x \in \mathbb{F}_{p^m}}\right) : \,\, a, b \in \mathbb{F}_{p^m}, c \in \mathbb{F}_p\right\}.
\end{split}
\end{equation}
The following result is useful for us to determine the weight distribution of $\mathcal{C}_f$, which is  given by \cite{Li2009}.
\begin{lemma}\cite[Theorem 2]{Li2009}\label{tlwm2}
Let $\bar{\mathcal{C}}_f$ be defined in (\ref{rwrwsf}) and $f(x)=x^{\frac{3^{k}+1}{2}}$, where $ 2\nmid k$ and gcd$(m,k)=1$.
\begin{description}
\item{\rm (1)} If $m$ is odd, then $\bar{\mathcal{C}}_f$ is a $[3^m,2m+1,2\cdot 3^{m-1}-3^{\frac{m-1}{2}}]$ code with weight distribution as follows:
    \begin{equation*}
\begin{split}
\left\{ \begin{array}{llll}
               0,   & \text{ occur} & 1 & \text{ time,}\\
            2\cdot 3^{m-1},  & \text{ occur} & (3^{m}-1)(3^m+3) & \text{ times,}\\
               2\cdot 3^{m-1}\pm 3^{\frac{m-1}{2}},  & \text{ occur} &  (3^{m}-1)3^m & \text{ times,}\\
            3^m, & \text{ occur}   & 2 & \text{ times.}\\
              \end{array}  \right.\ \\
\end{split}
\end{equation*}
\item{\rm (2)} If $m$ is even, then $\bar{\mathcal{C}}_f$ is a $[3^m,2m+1,2(3^{m-1}-3^{\frac{m-2}{2}})]$ code with weight distribution as follows:
\begin{equation*}
\begin{split}
\left\{ \begin{array}{llll}
               0,   & \text{ occur} & 1 & \text{ time,}\\
            2\cdot 3^{m-1},  & \text{ occur} & 3^{m+1}-3 & \text{ times,}\\
              2( 3^{m-1}\pm  3^{\frac{m-2}{2}}),  & \text{ occur} &  \frac{1}{2}(3^{m}-1)3^{m} & \text{ times,}\\
             2\cdot 3^{m-1}\pm 3^{\frac{m-2}{2}},  & \text{ occur} &  (3^{m}-1)3^{m} & \text{ times,}\\
            3^m, & \text{ occur}   & 2 & \text{ times.}\\
              \end{array}  \right.\ \\
\end{split}
\end{equation*}
\end{description}
\end{lemma}


With the above preparations, we now give the weight distribution of $\mathcal{C}_f$ for $f(x)=x^{\frac{3^{k}+1}{2}}$, where $ 2\nmid k$ and gcd$(m,k)=1$.
\begin{theorem}\label{Theorem 2}
Let $\mathcal{C}_f$ be the linear code defined in (\ref{code0}) and $f(x)=x^{\frac{3^{k}+1}{2}}$, where $ 2\nmid k$ and gcd$(m,k)=1$. Then the following statements hold.
\begin{description}
\item{\rm (1)} If $m$ is odd, then $\mathcal{C}_f$ is a $[3^m+1,2m+1,2\cdot3^{m-1}-3^{\frac{m-1}{2}}]$ code with weight distribution in Table \ref{Table5}. Its dual has the parameters $[3^m+1, 3^m-2m, 4]$.
\begin{table}[ht]
{\caption{\rm   The weight distribution of $\mathcal{C}_f$ for $m$ being odd}\label{Table5}
\begin{center}
\begin{tabular}{cccc}\hline
    $i$ & $A_i$ \\\hline
$0$  & $1$   \\
$3^m$  & $2$   \\
$2\cdot3^{m-1}$  & $3^{2m-1}+2\cdot3^m-3$   \\
$2\cdot3^{m-1}+1$  & $(3^m-3^{m-1})3^m$   \\
$2\cdot3^{m-1}\pm 3^{\frac{m-1}{2}}$  & $(3^{m-1}-1)3^m$   \\
$2\cdot3^{m-1}\pm 3^{\frac{m-1}{2}}+1$  & $(3^m-3^{m-1})3^m$   \\
  \hline
\end{tabular}
\end{center}}
\end{table}

\item{\rm (2)} If $m$ is even, then $\mathcal{C}_f$ is a $[3^m+1,2m+1,2\cdot3^{m-1}-2\cdot3^{\frac{m-2}{2}}]$ code with weight distribution in Table \ref{Table6}. Its dual has the parameters $[3^m+1, 3^m-2m, 4]$.
\begin{table}[h]
{\caption{\rm   The weight distribution of $\mathcal{C}_f$ for $m$ being even}\label{Table6}
\begin{center}
\begin{tabular}{cccc}\hline
     Weight & Multiplicity \\\hline
  $0$ & $1$  \\
 $3^m$ & $2$ \\
  $2\cdot3^{m-1}$ & $3(3^m-1)$ \\
  $2(3^{m-1}\pm3^{\frac{m-2}{2}})$ & $\left(3^{2m-1}-3^m\right)/2\mp3^{\frac{3m-2}{2}}$ \\
  $2(3^{m-1}\pm3^{\frac{m-2}{2}})+1$ & $3^{2m-1}\pm 3^{\frac{3m-2}{2}}$ \\
  $2\cdot3^{m-1}\pm3^{\frac{m-2}{2}}$ & $3^{2m-1}-3^m\pm2\cdot 3^{\frac{3m-2}{2}}$ \\
  $2\cdot3^{m-1}\pm3^{\frac{m-2}{2}}+1$ &  $2(3^{2m-1}\mp3^{\frac{3m-2}{2}})$\\  \hline
\end{tabular}
\end{center}}
\end{table}

\end{description}
\end{theorem}
{\it Proof.} Let $\mathbf{c}_f$ be a codeword in $\mathcal{C}_f$. It is easy to see that when $a=0$ and $(b,c)$ runs over $(\mathbb{F}_{3^m}, \mathbb{F}_3)$, the number of $\wt(\mathbf{c}_f)$ being $3^m$ or $3^{m-1}$ is $2$ or $3^{m+1}-3$, respectively. We only prove the weight distribution of $\mathcal{C}_{f}$ for $m$ being even. The case for $m$ being odd can be derived similarly.

From Lemma \ref{tlwm2}, we know that nonzero weights of $\mathcal{C}_f$ for $a\neq 0$ are as follows:
$$2( 3^{m-1}\pm 3^{\frac{m-2}{2}}),\,\,2( 3^{m-1}\pm 3^{\frac{m-2}{2}})+1,\,\,2\cdot 3^{m-1}\pm3^{\frac{m-2}{2}},\,\,2\cdot 3^{m-1}\pm3^{\frac{m-2}{2}}+1.$$
Let $w_1=2( 3^{m-1}+ 3^{\frac{m-2}{2}})$, $w_2=2( 3^{m-1}- 3^{\frac{m-2}{2}})$, $w_3=2( 3^{m-1}+ 3^{\frac{m-2}{2}})+1$, $w_4=2( 3^{m-1}- 3^{\frac{m-2}{2}})+1$, $w_5=2\cdot 3^{m-1}+3^{\frac{m-2}{2}}$, $w_6=2\cdot 3^{m-1}-3^{\frac{m-2}{2}}$, $w_7=2\cdot 3^{m-1}+3^{\frac{m-2}{2}}+1$ and $w_8=2\cdot 3^{m-1}-3^{\frac{m-2}{2}}+1$.
Let $A_{w_i}$ denote the number of codewords with  weight~$w_i$ for $1\leq i\leq 8$. From the representation of the codeword in $\mathcal{C}_f$ and Lemma~\ref{tlwm2}, we have
\begin{equation}\label{eqwdd}
\begin{split}
 \left\{ \begin{array}{llll}
A_{w_1}+A_{w_3}=A_{w_2}+A_{w_4}=\frac{1}{2}(3^m-1)3^m,\\
A_{w_5}+A_{w_7}=A_{w_6}+A_{w_8}=(3^m-1)3^m.
              \end{array}  \right.\ \\
\end{split}
\end{equation}
Moreover, the first five Pless power-moment
identities lead to the following system of equations:
\begin{equation}\label{eqwdd1}
\begin{split}
\begin{cases}
\sum_{i=1}^8w_iA_{w_i}=2\cdot3^{2m}\cdot (3^m+1)-2 \cdot 3^m-(3^{m+1}-3)\cdot 2\cdot 3^{m-1},\\
\sum_{i=1}^8w_i^2A_{w_i}=2\cdot3^{2m-1}(3^m+1)(2\cdot (3^m+1)+1)-2(3^m)^2-(3^{m+1}-3)(2\cdot 3^{m-1})^2,\\
\sum_{i=1}^8w_i^3A_{w_i}=2^3\cdot 3^{5m-2}+2^2\cdot 3^{4m}+46\cdot 3^{3m-2}+2\cdot 3^{2m}-2(3^m)^3-(3^{m+1}-3)(2\cdot 3^{m-1})^3,\\
\sum_{i=1}^8w_i^2A_{w_i}=16\cdot 3^{6m-3}+112 \cdot 3^{5m-3}+236\cdot 3^{4m-3}+226\cdot 3^{3m-3}+2\cdot 3^{2m}-2(3^m)^4-(3^{m+1}-3)(2\cdot 3^{m-1})^4.\\
\end{cases}
\end{split}
\end{equation}
From the equations (\ref{eqwdd}) and (\ref{eqwdd1}), we get $A_{w_1}=(3^{2m-1}-3^m)/2-3^{\frac{3m-2}{2}}$, $A_{w_2}=(3^{2m-1}+3^m)/2-3^{\frac{3m-2}{2}}$, $A_{w_3}=3^{2m-1}+ 3^{\frac{3m-2}{2}}$, $A_{w_4}=3^{2m-1}-3^{\frac{3m-2}{2}}$, $A_{w_5}=3^{2m-1}-3^m+2\cdot 3^{\frac{3m-2}{2}}$, $A_{w_6}=3^{2m-1}-3^m-2\cdot 3^{\frac{3m-2}{2}}$, $A_{w_7}=2(3^{2m-1}-3^{\frac{3m-2}{2}})$, $A_{w_8}=2(3^{2m-1}+3^{\frac{3m-2}{2}})$.

The parameters of the dual of $\mathcal{C}_f$ is easily obtained from Lemmas~\ref{lem:a1230} and \ref{lem:a4}. This completes the proof.  $\square$

\begin{example}\label{example3}
Let $\mathcal{C}_f$ be the linear code in Theorem \ref{Theorem 2}.
\begin{description}
\item{(1)} If $p=3,$ $m=2$, $f(x)=x^2$, then $\mathcal{C}_f $ has parameters $[10,5,4]$ and weight enumerator $1+18x^{4}+18x^{5}+96x^{6}+36x^{7}+36x^{8}+38x^{9}$.
 Its dual has parameters $[10,5,4]$.
  \item{(2)} If $p=3,$ $m=3$, $f(x)=x^2$,  then $\mathcal{C}_f$ has parameters $[28,7,15]$ and weight enumerator $1+216x^{15}+486x^{16}+294x^{18}+486x^{19}+216x^{21}+486x^{22}+2x^{27}$.
 Its dual has parameters $[28,21,4]$.
\item{(3)} If $p=3,$ $m=4$, $f(x)=x^{14}$,  then $\mathcal{C}_f $ has parameters $[82,9,48]$ and weight enumerator $1+1296x^{48}+1944x^{49}+1620x^{51}+4860x^{52}+240x^{54}+2592x^{57}+3888x^{58}+60x^{810}+2430x^{6}+2x^{81}$. Its dual has parameters $[82,73,4]$.
\end{description}
All of these codes and their duals are optimal or almost optimal with respect to the code tables in \cite{MGrassl}. These results have been verified by Magma.
\end{example}

\section{Conclusion }

This paper studied the linear code $\mathcal{C}_f$ given in~(\ref{code0}), which is a subfield code of the linear code from a known PN function~$f(x)$ and generalized some results in \cite{Hengar, Wang2020}. Some linear codes presented in this paper are optimal or almost optimal. Specifically, the main work is summarized as follows:

\begin{description}
\item{$\bullet$} In Section~$3$ and Section~$4$, we obtained the weight distribution of the subfield code $\mathcal{C}_f$ for $f(x)$ being a PN-DO function and a Coulter-Matthews function, respectively, and determined the parameters of the dual of $\mathcal{C}_f$.

\item{$\bullet$} Lemma~\ref{lem:detrelation} shows a relation between determinants of two related quadratic forms. This general result maybe helpful for application of
quadratic forms.

\item{$\bullet$} In Theorem \ref{Theorem 1}, the dual of $\mathcal{C}_f$ is a $p$-ary MDS code with parameters $[p+1,p-2,4]$ if $m=1$.

\item{$\bullet$} Example \ref{example1}, Example \ref{example2} and Example \ref{example3} showed  some optimal or almost optimal codes with respect to the code tables in \cite{MGrassl}.
\end{description}

\end{document}